\documentclass[osajnl,twocolumn,showpacs]{revtex4} 

\usepackage{epsfig}

\usepackage{times}

\begin{document}

\title{Blind dispersion compensation for optical coherence tomography}

\author{Konrad Banaszek, Aleksandr S. Radunsky, and Ian A. Walmsley}

\affiliation{Clarendon Laboratory, University of Oxford, Parks Road,
Oxford OX1 3PU, United Kingdom}

\date{\today}

\begin{abstract}
We propose a numerical method for compensating dispersion effects in optical coherence tomography that does not require {\em a priori} knowledge of dispersive properties of the sample. The method is based on the generalized autoconvolution function, and its principle of operation can be intuitively visualized using the Wigner distribution function formalism.
\end{abstract}


\maketitle

Since its inception about two decades ago,\cite{HuanSwanSCI91} optical coherence tomography (OCT) has evolved into an increasingly effective and promising imaging tool, particularly in biological and biomedical applications.\cite{OCTReview} The depth resolution of OCT is  proportional to the spectral bandwidth of the light source as long as optical dispersion of the medium can be neglected. However, for increasingly broadband sources employed to achieve ultrahigh axial resolution\cite{DrexJBO04} dispersion becomes relevant, broadening the coherence length of the measured signal and thus leading to the loss of resolution. 

Generally, two strategies have been developed to combat dispersion effects in OCT. 
The first one consists in matching experimentally the amount of dispersion in both the arms of the interferometer either with a simple introduction of an appropriate compensation plate in the reference arm,\cite{HitzBaumJBO99} or by a skilful design of the  scanning arrangement for the reference beam.\cite{TearBoumOL97,SmitZvyaOL02} The second strategy is to postprocess numerically complete fringe-resolved interferograms to compensate for the dispersion-induced loss of resolution.\cite{deBoSaxeAO01,FercHitzOC02,MarkOldeAO03} The numerical approach requires usually {\em a priori} knowledge of the dispersive properties of the medium which enter as the parameters of the compensation algorithm. 

In this paper we propose a numerical method for reconstructing dispersion-compensated depth profiles that does not involve knowledge of the exact dispersive characteristics of the medium. This method utilizes the phase information contained in complete complex interferograms to remove deleterious effects of dispersion. As we discuss later, it is equivalent to the recently demonstrated quantum OCT\cite{AbouNasrPRA02,NasrSalePRL03} which uses dispersion-cancellation effect in two-photon interference. Our method offers all the advantages of quantum OCT, but avoids the need for non-classical light sources. The dispersion compensation is performed by postprocessing data collected in a standard OCT setup with a low-coherence light using a simple and straightforward in implementation numerical algorithm. 

Before passing on to the detailed discussion of the method, we will present its principle of operation using a simple example. The input for our method is the complex envelope $\Gamma(\tau)$ of the full analytical mutual coherence function,\cite{deBoSaxeAO01,FercHitzOC02,MarkOldeAO03} parameterized with the delay of the reference beam. The dispersion-compensated depth profile is obtained by calculating the generalized autoconvolution of $\Gamma(\tau)$, defined as:
\begin{equation}
\label{Eq:Xiwtau}
\Xi_w(\tau) = \int \text{d}\tau' e^{-2w^2 \tau'^2} \Gamma^\ast(\tau+\tau')
\Gamma(\tau-\tau').
\end{equation}
The parameter $w$, not related directly to the dispersion characteristics of the medium, is used to tune the performance of the method, and it will be typically much smaller than the light bandwidth.

\begin{figure}

\epsfig{file=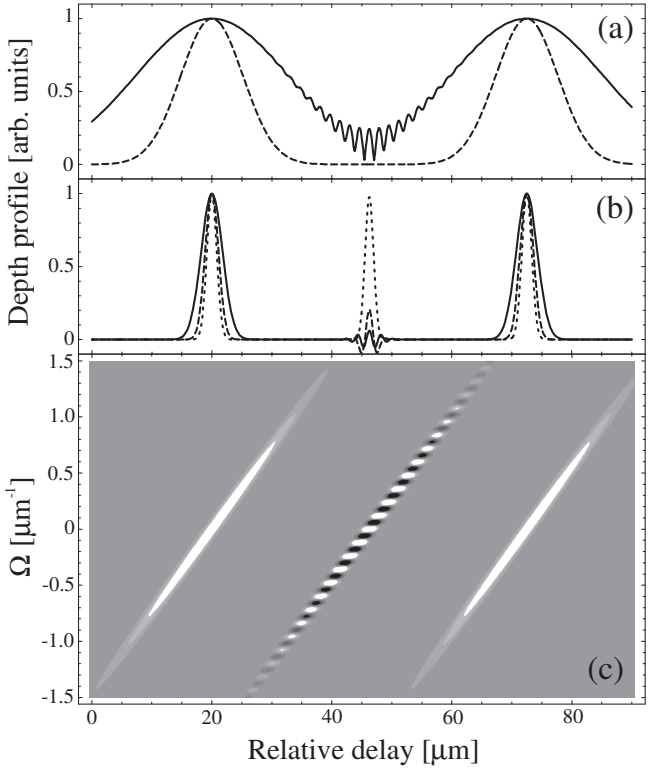,width=3.375in}

\caption{Reconstruction of a depth-profile of a pair of equally reflective surfaces preceded by 2~cm of dispersive aqueous region with group velocity dispersion $15$~fs$^{2}/$mm: (a) standard interferograms $|\Gamma(\tau)|$ for a coherence time $T=1.5~\mu$m (solid line) and $T=5.2~\mu$m (dashed line), the latter one giving optimal resultion with no dispersion compensation; (b) autoconvolution function $\Xi_w(\tau)$ for $w=0.015~\mu\mbox{m}^{-1}$ (dotted line), $w=0.06~\mu\mbox{m}^{-1}$ (dashed line), and $w=0.12~\mu\mbox{m}^{-1}$ (solid line); (c) the chronocyclic Wigner function $W(\tau,\Omega)$ of the complex interferogram envelope $\Gamma(\tau)$. For convenience, time and frequency have been expressed in length units using the vacuum speed of light. The profiles in graphs (a) and (b) have been renormalized to the same height in order to make the comparison of their widths easier.}
\label{Fig:TwoSurfaces}
\end{figure}

The autoconvolution function defined above can reveal the axial structure of a sample with a better resolution than the interferogram itself. We illustrate this with Fig.~\ref{Fig:TwoSurfaces}(a,b), where we present a reconstruction of an exemplary depth profile for a pair of reflective surfaces embedded in a dispersive material and illuminated with a broadband light. First, let us consider the scenario when no attempt to compensate for dispersion effects has been made. Fig.~\ref{Fig:TwoSurfaces}(a) shows the absolute value of the envelope $|\Gamma(\tau)|$ which is the most basic way to retrieve the depth profile. This profile is severely broadened by dispersion. For comparison we also plot in this graph a profile that would be obtained for a narrower bandwidth of the probe light, giving the optimal resolution of $|\Gamma(\tau)|$ for this specific medium. These two plots should be contrasted with Fig.~\ref{Fig:TwoSurfaces}(b) which shows the autoconvolution function calculated according to Eq.~(\ref{Eq:Xiwtau}) for several values of the parameter $w$. It is seen that the autoconvolution function reveals the location of the reflective surfaces as two sharp peaks with a resolution comparable to the coherence length of the probe light itself. Additionally, the convolution function contains a spurious artefact located half-way between the peaks. Its presence is a result of numerical interference of signals reflected by the two surfaces, but its magnitude can be quickly suppressed by increasing the value of the parameter $w$. The suppression is accompanied by a slight broadening of the genuine peaks in the depth profile, which is however significantly less severe than that affecting $|\Gamma(\tau)|$. 

Let us now discuss in detail properties of the autoconvolution function $\Xi_w(\tau)$ as a tool for reconstructing depth profiles. We consider a standard OCT setup in which broadband light, characterized by the power spectrum $S(\omega)$ centered around a frequency $\omega_0$,
is split into two beams. One, signal beam is reflected off the sample, thus acquiring in the spectral domain the response function of the sample $\varrho(\omega)$, whereas the second reference beam undergoes a controlled temporal delay $2\tau$. Interference between these two beams yields the mutual coherence function, given by:
\begin{equation}
\Gamma(\tau)e^{-2i\omega_0\tau}
=  
\int \text{d}\omega \, S(\omega) \varrho(\omega) e^{-2i\omega \tau}.
\end{equation}
In this formula, we separated out the phase factor rapidly oscillating with the optical frequency $\omega_0$, and denoted the resulting slowly-varying complex envelope as $\Gamma(\tau)$.

We will model the spectral response function $\varrho(\omega)$ as composed of discrete contributions coming from reflective surfaces within the sample characterized by reflection coefficients $r_n$:
$
\varrho(\omega) = \sum_{n} r_n e^{2i\varphi_n(\omega)}$.
The phase $\varphi_n$ acquired by the signal field reflected from the $n$th surface can be expanded around the central frequency of the probe light up to the quadratic term:
\begin{equation}
\varphi_n(\omega) \approx \varphi_n(\omega_0) + (\omega-\omega_0) \tau_n
+ \frac{1}{2} (\omega-\omega_0)^2 D_n + \ldots
\end{equation}
We will incorporate the constant phase $\varphi_n(\omega_0)$ into the reflection coefficient $r_n$. The parameter $\tau_n$ multiplying the linear term characterizes the position of the $n$th reflective surface, whereas $D_n$ describes
dispersion affecting the component reflected from that surface.
We will also assume the Gaussian spectrum of the probe light
$S(\omega) \propto \exp(-T^2(\omega - \omega_0)^2)$
with $T$ characterizing its coherence time. Throughout this paper, we will use half-width at $1/e$-maximum as a measure of the resolution. Within the introduced model, the complex envelope of the mutual coherence function is a sum of contributions from the surfaces $\Gamma(\tau)=\sum_n \Gamma_n(\tau)$ given by:
\begin{equation}
\label{Eq:Gammantau}
\Gamma_n(\tau) \propto r_n
\exp\left(-\frac{(\tau-\tau_n)^2}{T^2-iD_n} \right)
\end{equation}
In standard OCT, the depth profile is retrieved directly from the interferogram as the absolute value $|\Gamma_n(\tau)|$, and consequently the $n$th surface is visualized as a peak with a dispersion-broadened width $\sqrt{T^2+D_n^2/T^2}$.

Let us now turn to the analysis of the information on the depth profile contained in the generalized autoconvolution function $\Xi_w(\tau)$. With the complex envelope $\Gamma(\tau)$ given as a sum of terms calculated in Eq.~(\ref{Eq:Gammantau}), the generalized autoconvolution function can be decomposed into a double sum $\Xi_w(\tau)=\sum_{mn}\Xi_w^{(mn)}(\tau)$ of contributions obtained by inserting a product $\Gamma^\ast_m(\tau+\tau')\Gamma_n(\tau-\tau')$ into Eq.~(\ref{Eq:Xiwtau}).
We will analyze separately the diagonal terms with $m=n$, which as we will see reveal positions of the reflective surfaces, and then the cross-terms  $\Xi_w^{(mn)}(\tau)$ with $m\neq n$ that are responsible for the artefacts in the reconstructed profile, like the one we have seen in Fig.~\ref{Fig:TwoSurfaces}(b).
The complete analytical expressions are rather complicated and we will approximate them by performing an expansion up to the leading order of $w$. This will give us an insight into relative scales of parameters involved in the procedure.

For $m=n$, an explicit calculation yields the following expression for $\Xi_w^{(nn)}(\tau)$ in the limit when $w^2 \ll 1/(T^2 + D_n^2/T^2)$:
\begin{equation}
\label{Eq:Xiwnntau}
\Xi_w^{(nn)}(\tau) \propto |r_n|^2 \exp
\left(-\frac{2(\tau-\tau_n)^2}{T^2 + (D_n w)^2}
\right)
\end{equation}
This expression describes a Gaussian peak located at the position $\tau_n$ of the $n$th reflective surface. The width of this peak is given by $\sqrt{[T^2+(D_n w)^2]/2}$, and in the limit when $w\rightarrow 0$ it approaches the dispersion-free limit defined solely by the coherence time of the light source, equal to $T/\sqrt{2}$. Compared to the standard interferogram envelope $|\Gamma(\tau)|$ in the absence of dispersion, the peak in the autoconvolution function is narrower by a factor $\sqrt{2}$; this narrowing is easily understandable as $\Xi_w(\tau)$ is quadratic in $\Gamma(\tau)$. 

The purpose of introducing the parameter $w$ is to suppress the cross-terms $\Xi_w^{(mn)}(\tau)$ with $m\neq n$. In order to keep the interpretation of the mathematical expressions simple, we will restrict our attention to the regime when dispersion affecting contributions from two reflecting surfaces is comparable, i.e.\ $D_m \approx D_n \approx \bar{D}$. This is the case when the bulk of dispersion comes from the medium preceding both the surfaces. In this regime, it is possible to give a simple formula for the magnitude of the cross-terms:
\begin{eqnarray}
|\Xi_w^{(mn)}(\tau)| & \propto & |r_m^\ast r_n | \exp
\left(
-\frac{2(\tau-\bar{\tau})^2}{T^2 + (\bar{D} w)^2}
\right)
\nonumber
\\
\label{Eq:Ximntau}
& & \times \exp\left( -\frac{w^2}{2}(\tau_m - \tau_n)^2 \right)
\end{eqnarray}
where the proportionality factor is the same as in Eq.~(\ref{Eq:Xiwnntau}) and $\bar{\tau}=(\tau_m+\tau_n)/2$. This formula describes a structure located half-way between the positions of the contributing surfaces. The magnitude of the structure is a function of $w$ through the multiplicative factor $\exp[-w^2(\tau_m-\tau_n)^2/2]$. Its exponential dependence on $w$ allows us to suppress efficiently the spurious cross-terms in the autoconvolution function by setting a non-zero value of $w$, with only a slight worsening of the resolution exhibited in Eq.~(\ref{Eq:Xiwnntau}). The supression of the cross-terms requires $|w|$ exceeding $1/|\tau_m-\tau_n|$ and therefore is more efficient for larger separation between the peaks.

The operation of the blind dispersion compensation method can be understood intuitively with the help of the chronocyclic Wigner distribution function,\cite{WignerReview} defined for the complex interferogram envelope $\Gamma(\tau)$ in the standard way as:
\begin{equation}
W(\tau,\Omega) = \frac{1}{\pi} \int \text{d}\tau' \, e^{-2i\Omega\tau'}
\Gamma^{\ast}(\tau+\tau') \Gamma(\tau-\tau')
\end{equation}
In Fig.~\ref{Fig:TwoSurfaces}(c) we depict the Wigner function for the example discussed in this paper. The Wigner function contains two peaks corresponding to the two reflective surfaces, and an oscillating interference pattern located half-way between the peaks. 
This pattern is a signature of coherence between the two reflections. Dispersion introduces time-frequency correlations which result in a tilt clearly seen in Fig.~\ref{Fig:TwoSurfaces}(c). In standard OCT the depth profile is retrieved as $|\Gamma(\tau)|$ which is given as a square root of $W(\tau,\Omega)$ integrated along the $\Omega$ axis. Then the dispersion-induced tilt severely deteriorates the resolution. However, taking a cross-section through the Wigner function along a horizontal line for $\Omega=0$, yields a dispersion-free profile. The problem with this profile is that it contains strong contributions from the interference pattern\cite{HlawFlanWIG97} which were washed out in $|\Gamma(\tau)|$ due to the integration over the frequency variable. The answer to this problem is to carefully average the profile over a range of frequencies $\Omega$. This is exactly the purpose of the generalized autoconvolution function in Eq.~(\ref{Eq:Xiwtau}), which can be rewritten in the Wigner formalism as:
\begin{equation}
\Xi_w(\tau)= \sqrt{\frac{\pi}{2w^2}}
\int \text{d}\Omega \,
 W(\tau,\Omega) e^{-\Omega^2/2w^2}.
\end{equation}
The averaging along the frequency axis over an interval defined by the parameter $w$ rapidly washes out the contribution from the interference pattern, while nearly retaining the width of the genuine peaks. The width of the interval must be larger than the spacing of the interference pattern, which in turn is inversely proportional to the separation between the peaks. As shown by Abouraddy {\em et al.}\cite{AbouNasrPRA02}, the effects of dispersion can be removed using a quantum effect of two-photon interference. Interestingly, the depth profile obtained from the joint detection of two photons in their case yields the same order correlation function as $\Xi_w(\tau)$. This equivalence is related to the fact that in the quantum OCT scheme only one of the two photons passes through the dispersive medium.

In conclusion, we have shown that the generalized autoconvolution function calculated from the complex interferogram envelope can reveal location of reflective surfaces in a dispersive medium with a resolution reaching the coherence length of the employed light itself. We expect that in the case of more complex depth profiles, the oscillatory character of the artefacts will lead to similar suppression efficiency as that in the simple numerical example discussed here.

\end{document}